\documentclass[reprint,aps,graphicx,pre,showpacs]{revtex4-1}
\usepackage{amsmath,amssymb,amsfonts}     

\usepackage{graphicx,bm}
  \usepackage{paralist}
  \usepackage{epstopdf}
  \usepackage{graphics} 
 \usepackage[colorlinks=true]{hyperref}
 \hypersetup{urlcolor=blue, citecolor=red}
 \usepackage[latin1]{inputenc}
\usepackage[caption=false]{subfig}

\renewcommand{\theequation}{\arabic{section}.\arabic{equation}}

\renewcommand{\d}{{\rm d}}

\newcommand{\R}{{\mathbb R}}

\newcommand{\e}{{\rm e}}

\newcommand{\calU}{{\mathcal U}}

\newcommand{\x}{\mathbf{x}}
\newcommand{\y}{\mathbf{y}}

\newcommand{\X}{\mathbf{X}}
\renewcommand{\P}{\mathbb{P}}

\newcommand{\calT}{\mathcal{T}}
\newcommand{\calM}{\mathcal{M}}
\newcommand{\calG}{\mathcal{G}}
\newcommand{\calQ}{\mathcal{Q}}

\newcommand{\n}{\mathbf n}

\renewcommand{\theequation}{\arabic{section}.\arabic{equation}}

\newcommand{\E}{{\mathbb E}}

\begin{document}
\title{Truncated stochastically switching processes}
\author{Paul C. Bressloff}
\address{Department of Mathematics, Imperial College London, London SW7 2AZ, UK.}

\date{\today}

\begin{abstract}
 There are a large variety of hybrid stochastic systems that couple a continuous process with some form of stochastic switching mechanism. In many cases the system switches between different discrete internal states according to a finite-state Markov chain, and the continuous dynamics depends on the current internal state. The resulting hybrid stochastic differential equation (hSDE) could describe the evolution of a neuron's membrane potential, the concentration of proteins synthesized by a gene network, or the position of an active particle. Another major class of switching system is a search process with stochastic resetting, where the position of a diffusing or active particle is reset to a fixed position at a random sequence of times. In this case the system switches between a search phase and a reset phase, where the latter may be instantaneous. In this paper, we investigate how the behavior of a stochastically switching system is modified when the maximum number of switching (or reset) events in a given time interval is fixed. This is motivated by the idea that each time the system switches there is an additive energy cost. We first show that in the case of an hSDE, restricting the number of switching events is equivalent to truncating a Volterra series expansion of the particle propagator. Such a truncation significantly modifies the moments of the resulting renormalized propagator. We then investigate how restricting the number of reset events affects the diffusive search for an absorbing target. In particular, truncating a Volterra series expansion of the survival probability, we calculate the splitting probabilities and conditional MFPTs for the particle to be absorbed by the target or to exceed a given number of resets, respectively.
\end{abstract}

\maketitle


\section{Introduction}

There are a wide range of stochastic processes in cell biology that involve the coupling between continuous and discrete random variables (stochastic hybrid systems) \cite{Bressloff17a}. The continuous process could represent the concentration of proteins synthesized by a gene \cite{Kepler01,Bose04,Smiley10,Newby12,Newby15,Hufton16}, the membrane voltage of a neuron \cite{Fox94,Chow96,Keener11,Goldwyn11,Buckwar11,NBK13,Bressloff14b,Newby14}, the position of a swimming bacterium \cite{Berg77,Hillen00,Berg04,Erban05}, or a molecular motor \cite{Reed90,Friedman05,Newby10,Bressloff11,Bressloff13}. The corresponding discrete process could represent the activation state of the gene, the conformational state of an ion channel, or the velocity state of an active particle. Let $(\X(t),N(t))$ denote the state of the system at time $t$ with $\X(t)\in \R^d$ and $N(t) \in \Gamma$, where $\Gamma$ is a discrete set. Assuming that $N(t)=n$, the continuous variables typically evolve according to a hybrid stochastic differential equation (hSDE) of the form  $ d\X ={\bf A}_n(\X)dt+\sqrt{2D}d{\bf W}$, where ${\bf W}$ is a vector of independent Wiener processes and ${\bf A}_n$ is an $n$-dependent drift term. (The diffusivity could also depend on $n$.) The discrete variable switches between the different discrete states according to a continuous time Markov chain whose matrix generator could itself depend on $\X(t)$. In the limit $D\rightarrow 0$, the dynamics reduces to a so-called piecewise deterministic Markov process \cite{Davis84}.

In many applications of hSDEs, there is a separation of time scales, whereby the switching between discrete states of the Markov chain is fast compared to the relaxation dynamics of the continuous process. Suppose that $\tau$ is the characteristic time-scale of the relaxation dynamics and $\epsilon\tau$ is the characteristic time-scale of the Markov chain for some small positive parameter $\epsilon$. Taking the limit $\epsilon\rightarrow 0$ then leads to an effective continuous dynamical system that is obtained by averaging the piecewise dynamics with respect to the corresponding unique stationary measure of the Markov chain (assuming the latter exists). In the weak-noise regime $0<\epsilon \ll 1$, various approaches have been used to study noise-induced transitions between metastable states of the averaged system. These include large deviation theory \cite{Kifer09,fagg09,fagg10,Bressloff17}, WKB approximations and matched asymptotics \cite{Keener11,Newby12,NBK13,Bressloff14b,Newby15}, and stochastic hybrid path integrals \cite{Bressloff14,Bressloff21a,Bressloff21b}.

Another important example of a randomly switching process is a search process with stochastic resetting. (See the review \cite{Evans20} and references therein.) The simplest version of a resetting protocol is to instantaneously reset the position of a diffusing particle to some fixed point $\x_r$ at a constant rate  $r$ \cite{Evans11a,Evans11b,Evans14}. One of the characteristic properties of a search process with stochastic resetting is that the mean time for a Brownian particle to find a hidden target in an unbounded domain is finite, and has an optimal value as a function of the resetting rate $r$. This is a consequence of the fact that the mean first passage time (MFPT) to find the target diverges in the limits $r\rightarrow 0$ and $r\rightarrow \infty$. Analogous behavior has been observed in other search processes with resetting, including non-diffusive search processes such as Levy flights \cite{Kus14}, active run and tumble particles \cite{Evans18,Bressloff20} and directed velocity jump processes \cite{Bressloff20d,Bressloff20e}, diffusion in potential landscapes \cite{Pal15,Pal19} or switching environments \cite{Bressloff20a,Bressloff20b,Mercado21}, resetting followed by a refractory period \cite{Evans19a,Mendez19a}, resetting with finite return times \cite{Pal19,Pal19a,Mendez19,Bodrova20,Pal20,Bressloff20c}, and encounter-based models of absorbing targets \cite{Bressloff22a,Bressloff22b,Grebenkov22}.

In this paper we consider a different aspect of stochastically switching systems, namely, conditioning the process on the maximum number of switching events that can occur. That is, if $\calM(t)$ denotes the number of switching events in the interval $[0,t]$, then we impose the condition $\calM(t)\leq \mu<\infty$ for all $t$. One motivation for such a construction is that state transitions in an hSDE tend to cost energy, so that the maximum number of such transitions could be limited. Alternatively, conditioning on the number of transitions provides another type of statistic that could be measured experimentally. For example, in the case of gene networks, transitions from the inactive to active state often results on some form of bursting. In the case of search processes, the cost of stochastic resetting has been explored in a recent paper \cite{Sunil23}, which assumes that the cost is additive, and the contribution of each reset is a function of the distance a particle must travel to the reset position $\x_r$. These authors focus on the mean cost accrued by a search process that is terminated when the target is found. In contrast, we take the cost to be equal to the number of reset events, and terminate the search process as soon as one or other of the following occurs: the particle finds the target or the number of reset events crosses some threshold.

The structure of the paper is as follows. In Sect. II we give a general definition of an hSDE  and write down the evolution equation for the associated propagator. In Sect. III we construct an integral equation for the propagator, which is expanded as a Volterra series, whose individual terms correspond to fixing the number of state transitions. Truncating the Volterra series is then equivalent to restricting the maximum number of allowed state transitions. We use this to define a renormalized propagator and its associated moments.
The theory is illustrated in Sect. IV using the example of an OU process with random drift, which has previously been used to model the motion of an RTP with diffusion in a harmonic potential \cite{Basu20,Garcia21} and protein synthesis in a two-state gene network \cite{Bose04,Smiley10}. We use the corresponding diagrammatic expansion to calculate moments of the hSDE that are conditioned on the maximum number of switching events. In Sect. V, we develop the analogous theory for a diffusive search process with stochastic resetting. In this case, we expand the standard last renewal equation for the survival probability as a Volterra series in the number of resetting events. Truncating the series now corresponds to restricting the maximum number of resets. We use this to calculate the splitting probabilities and conditional MFPTs for the particle to be absorbed by the target or to exceed a given number of resets, respectively.

\section{Hybrid SDE in $\R^d$}

Consider a system whose states are described by a pair
of stochastic variables $(\X(t),N(t)) \in \R^d \times \{0,\cdots,K-1\}$. When the discrete state is $N(t)=n$, the system evolves 
according to the SDE
\begin{equation}
\label{PDMP}
d\X(t)={\bf A}_{n}(\X(t))dt+\sqrt{2D}d{\bf W}(t),
 \end{equation}
where ${\bf W}$ is a vector of $d$ independent Wiener processes. The discrete stochastic variable $N(t)$ evolves according to a $K$-state
continuous-time Markov chain with a $K\times K$ matrix generator ${\bf Q}$ that is taken to be independent of $\X(t)$. It is related to the corresponding transition matrix ${\bf W}$ according to
\begin{equation}
Q_{nm} =W_{nm}-\delta_{n,m} \Gamma_m, \quad \Gamma_m = \sum_{k=0}^{K-1}W_{km}.
\end{equation}
Given the definition of $\Gamma_m$, we can introduce the decomposition $W_{nm}=P_{nm}\Gamma_m $ with $\sum_nP_{nm}=1$. The positive quantity $\Gamma_m$ is the rate at which a transition from the state $m$ occurs and $P_{nm}$ is the probability that such a transition is to the state $n$. We assume that the generator is irreducible so that there exists a stationary density ${\bm \rho}$ for which $\sum_mQ_{nm}\rho_m=0$. In the case of a two-state hSDE ($n=0,1$), the matrix generator takes the form
\begin{equation}
{\bf Q}(x)= \left (\begin{array}{cc} -\beta   &\quad \alpha 
\\ \beta & \quad -\alpha  \end{array}
\right ),
\label{Q}
\end{equation}
and
\begin{equation}
\rho_0=\frac{\alpha}{\alpha+\beta},\quad \rho_1=1-\rho_0=\frac{\beta}{\alpha+\beta}.
\end{equation}

Given the initial conditions $\X(0)=\x_0,N(0)=n_0$, we introduce the propagator $G_{nn_0}(\x,t|\x_0,0) $ with
\begin{align}
&G_{nn_0}(\x,t|\x_0,,0)d\x\nonumber \\
&=\P[\X(t)\in (\x,\x+d\x),\, N(t) =n|\x_0,n_0],
\end{align}
and $G_{nn_0}(x,0|x_0,0)=\delta_{n,n_0}\delta(x-x_0)$.
The propagator $G_{nn_0}$ evolves according to the forward differential Chapman-Kolmogorov (CK) equation 
\begin{align}
\label{CKH}
 \frac{\partial G_{nn_0}}{\partial t}&=-{\bm \nabla} \cdot [{\bf A}_n(\x)G_{nn_0}]+D{\bm {\bm \nabla}}^2G_{nn_0}\nonumber \\&\quad +\sum_{m=0}^{K-1}Q_{nm} G_{mn_0}.
\end{align}
The first two terms on the right-hand side represent the probability flow associated with the SDE for a given $n$, whereas the third term 
represents jumps into or out of the discrete state $n$. In the absence of switching with $n$ fixed, the system reduces to a single SDE whose corresponding FP equation takes the form
\begin{align}
\label{FP}
 \frac{\partial p_{n}}{\partial t}&=-{\bm \nabla} \cdot [{\bf A}_n(\x)p_{n}]+D{\bm {\bm \nabla}}^2p_{n},
 \end{align}
 and $p_n(\x,0|\x_0,0)=\delta(\x-\x_0)$. We will refer to $p_n$ as the bare (no switching) propagator.

\setcounter{equation}{0}
\section{Integral equation and Volterra series expansion}

The propagator $G_{nm}$ satisfies an integral equation of the form
\begin{widetext}
\begin{align}
\label{renewal0}
&G_{nm}(\x,t|\x_0,0)=\delta_{n,m}\e^{-\Gamma_{m} t}p_{m}(\x,t|\x_0,0)+\sum_{l} W_{nl} \int_0^{t}d\tau d\y\, \e^{-\Gamma_{n} (t-\tau)}p_{n}(\x,t|\y,\tau)G_{lm}(\y,\tau|\x_0,0).
\end{align}
The first term on the right-hand side is the contribution from all paths that never switch in the interval $[0,t]$, which only occurs if $n=m$. The probability of no switching from the state $m$ is $\e^{-\Gamma_m t}$. The second term on the right-hand side represents the sum over all trajectories that switch at least once, with the final transition occurring at the time $\tau$. 
Iterating the integral Eq. (\ref{renewal0}) generates a Volterra series representation of the propagator:
\begin{align}
\label{renewalV}
&G_{nm}(\x,t|\x_0,0)=\delta_{n,m}\e^{-\Gamma_{m} t}p_{m}(\x,t|\x_0,0)+ W_{nm}\int_0^{t}d\tau \int d\y\  \e^{-\Gamma_{n} (t-\tau) }p_{n}(\x,t|\y,\tau)\e^{-\Gamma_{m} \tau}p_{m}(\y,\tau|\x_0,0)\nonumber \\ 
& \ + \sum_l W_{nl}W_{lm} \int_0^{t}d\tau_2 \int_0^{\tau_2}d\tau_1\int d\y_2\int d\y_1\, \e^{-\Gamma_{n} (t-\tau_2) }p_{n}(\x,t|\y_2,\tau_2)\e^{-\Gamma_{l} (\tau_2-\tau_1)}p_{l}(\y_2,\tau_2|\y_1,\tau_1)\e^{-\Gamma_{m}  \tau_1}p_{m}(\y_1,\tau_1|\x_0,0)\nonumber \\
&\quad +\ldots
\end{align}
\end{widetext}
 The $j$th term in the series expansion, $j \geq 0$, has the following interpretation: it specifies the contribution to the propagator from paths that undergo exactly $j$ switching events. For example, if $n\neq m$ then the $j=1$ term has a factor $P_{nm}\Gamma_m \e^{-\Gamma_n(t-\tau)}\e^{-\Gamma_m\tau}$, after setting $W_{nm}=P_{nm}\Gamma_m$. The probability that the first transition occurs in the time interval $[\tau,\tau+d\tau]$ is $\Gamma_m \e^{-\Gamma_m\tau}d\tau$, the probability that $m\rightarrow n$ is $P_{nm}$, and the probability that there are no transitions from the state $n$ is $\e^{-\Gamma_n(t-\tau)}$. Hence, the total probability that there is a single transition $m\rightarrow n$ in the time interval $[0,t]$ is
 \begin{align}
 P^{(1)}_{nm}(t)&=W_{nm}  \e^{-\Gamma_nt}\int_0^t \e^{-[\Gamma_m-\Gamma_n]\tau}d\tau\nonumber \\
 &=\frac{W_{nm}}{\Gamma_m-\Gamma_n}\left [ \e^{-\Gamma_nt}-\e^{-\Gamma_mt}\right ].
 \end{align}
 Similarly, the probability that there are two transitions in the interval $[0,t]$ is
 \begin{widetext}
  \begin{align}
 P^{(2)}_{nm}(t)&=\sum_lW_{nl}W_{lm}\int_0^{t}d\tau_2  \int_0^{\tau_2}d\tau_1 \e^{-\Gamma_{n} (t-\tau_2) } \e^{-\Gamma_{l} (\tau_2-\tau_1)}\e^{-\Gamma_{m}  \tau_1} =\sum_l\frac{W_{nl}W_{lm}}{\Gamma_m-\Gamma_l}\left [\frac{\e^{-\Gamma_nt}-\e^{-\Gamma_lt}}{\Gamma_l-\Gamma_n}-\frac{\e^{-\Gamma_nt}-\e^{-\Gamma_mt}}{\Gamma_m-\Gamma_n}\right ]. 
 \end{align}
 \end{widetext}
 In addition, integrating Eq. (\ref{renewalV}) with respect to $\x$, summing over $n$, and then using the unit normalization of the propagator shows that
 \begin{equation}
 1=\e^{-\Gamma_m t} +\sum_{j\geq 1}\sum_n P^{(j)}_{nm}(t).
 \end{equation}

 \begin{figure*}[t!]
\centering
\includegraphics[width=17cm]{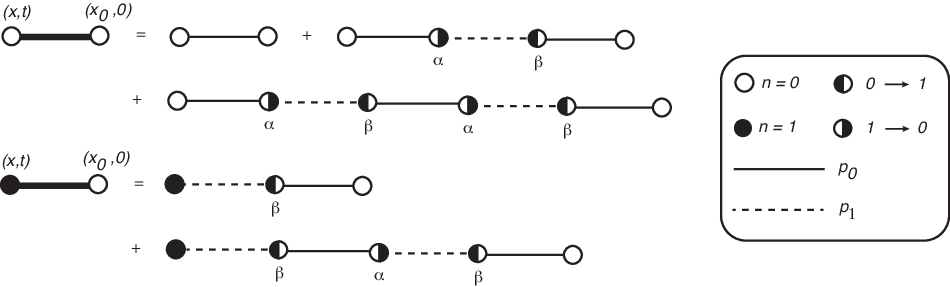} 
\caption{First few terms in the diagrammatic expansions of the full propagators $G_{00}(\x,t|\x_0,0)$ and $G_{10}(\x,t|\x_0,0)$ in terms of the bare propagators $p_n(\x,t|\x_0,t_0)$ for the two-state Markov chain. Time flows from right to left.}
\label{fig1}
\end{figure*}
 
For the sake of illustration, consider a two-state hSDE with matrix generator (\ref{Q}). Suppose that the system starts in the state $n_0=0$. Then
\begin{subequations}
\label{renewal}
\begin{align}
& G_{00}(\x,t|\x_0,0)=\e^{-\beta t}p_{0}(\x,t|\x_0,0) \\
& \qquad +\alpha \int_0^{\infty}d\tau  d\y\, \e^{-\beta (t-\tau)} p_{0}(\x,t|\y,\tau)G_{10}(\y,\tau|\x_0,0), \nonumber \\
&G_{10}(\x,t|\x_0,0)\\
&\quad =\beta \int_0^{\infty}d\tau \int d\y\, \e^{-\alpha (t-\tau)}p_{1}(\x,t|\y,\tau)G_{00}(\y,\tau|\x_0,0).\nonumber
\end{align}
Similarly, if $n_0=1$, then
\begin{align}
& G_{01}(\x,t|\x_0,0)\\
&\quad =\alpha \int_0^{\infty}d\tau\int d\y  \,  \e^{-\beta (t-\tau)}  p_{0}(\x,t|\y,\tau)G_{11}(\y,\tau|\x_0,0) \nonumber \\
&G_{11}(\x,t|\x_0,0)=\e^{-\alpha t}p_{1}(\x,t|\x_0,0) \\
& \qquad +\beta  \int_0^{\infty}d\tau  d\y\, \e^{-\alpha[t- \tau]}p_{1}(\x,t|\y,\tau)G_{01}(\y,\tau|\x_0,0).\nonumber
\end{align}
\end{subequations}
The first term on the right-hand side of Eq. (\ref{renewal}a) represents the contribution from all trajectories that never switch to the state $n=1$. The latter occurs with probability $\e^{-\beta t}$. On the other hand, the integral term sums over all trajectories that switch at least once, with the last switch $1\rightarrow 0$ occurring at a rate $\beta$ at a time $\tau$, $0<\tau<t$. Similar interpretations apply to Eqs. (\ref{renewal}b-d). Iterating Eq. (\ref{renewal}a) gives
\begin{align}
\label{tr2}
 &G_{00}(\x,t|\x_0,0)\\
 &=\e^{-\beta t}p_{0}(\x,t|\x_0,0)+ \alpha \beta\int_{0}^td\tau_2 \int_{0}^{\tau_2}d\tau_1 \int d\y_2\int d\y_1 \nonumber \\
 &\times \e^{-\beta (t-\tau_1)}  p_0(\x,t|\y_2,\tau_2)\e^{-\alpha(\tau_2-\tau_1)}p_1(\y_2,\tau_2|\y_1,\tau_1)\nonumber \\
 &\times\e^{-\beta \tau_2}p_0(\y_1,\tau12|\x_0,0)+\ldots.\nonumber
\end{align}
Since, the initial and final discrete states are the same, the number of switches has to be even.
 Using similar arguments, we obtain analogous series expansions of $G_{11}, G_{01}$ and $G_{10}$. For example, contributions to $G_{11}$ involve sequences of transitions of the form $1\rightarrow 0 \rightarrow 1$, whereas contributions to $G_{10}$ involves the transition $1\rightarrow 0$ followed by additional transitions of the form $0\rightarrow 1\rightarrow 0$. The first few terms in the diagrammatic expansions of $G_{00}$ and $G_{10}$ are shown in Fig. \ref{fig1}.

A few comments are in order. First, as we show in section IV, the series expansion (\ref{tr2}) is not uniformly convergent due to the presence of secular terms involving powers of $\alpha t$ and $\beta t$. Thus one cannot interpret Eq. (\ref{tr2}) as a perturbation expansion in the slow switching limit $\alpha,\beta\rightarrow 0$. On the other hand, as we have already highlighted, the terms in Eq. (\ref{tr2}) have a natural probabilistic interpretation based on the number of state transitions. In particular, truncating the series is equivalent to conditioning the propagator with respect to the maximum number of transitions. For a general hSDE, let $G^{(\mu)}_{nm}(\x,t|\x_0,0)$ denote the contribution to the propagator from paths that have a maximum of $\mu$ transitions, which is given by the first $\mu+1$ terms in the corresponding diagrammatic expansion. Taking the random variable ${\mathcal M}(t)$ to denote the number of transitions over the interval $[0,t]$, it follows that
\begin{align}
\P[\calM(t) \leq \mu]&=\int d\x \, {G}^{(\mu)}_{nm}(\x,t|\x_0,0)\nonumber \\
&=\delta_{n,m}\e^{-\Gamma_n t} +\sum_{j=1}^{\mu}P^{(j)}_{nm}(t).
\end{align}
We then introduce a renormalized propagator that is conditioned to undergo a maximum of $\mu$ transitions:
\begin{align}
\calG^{(\mu)}_{nm}(\x,t|\x_0,0)
&=\frac{G^{(\mu)}_{nm}(\x,t|\x_0,0)}{\sum_{l=0}^{K-1}\int_{-\infty}^{\infty} G^{(\mu)}_{lm}(\x,t|\x_0,0)d\x}.
\label{rho}
\end{align}

\setcounter{equation}{0}
\section{OU process with random drift}

In this section we illustrate the theory by considering the particular example of an OU process with random drift. This has previously been used to model an RTP with diffusion in a harmonic potential \cite{Basu20,Garcia21} and protein synthesis in a gene network \cite{Bose04,Smiley10}. In the former case, $X(t)\in \R$ represents the position of the RTP at time $t$ whereas $N(t)=n\in \{0,1\}$ specifies the current velocity state $v_n$ of the particle. If $v_0=v$ and $v_1=-v$ then the motion becomes unbiased when the mean time spent in each velocity state is the same ($\alpha=\beta$). On the other hand, in the case of the gene network, $X(t)$ represents the concentration of synthesized protein and $N(t)$ specifies whether the gene is active or inactive. That is, $v_n$ is the rate of synthesis with $v_0>v_1\geq 0$. In both examples, the variable $X(t)$ evolves according to the piecewise SDE
\begin{equation}
\label{PDMPrtp}
dX(t)=[-\kappa_0X(t)+v_n]dt+\sqrt{2D}dW(t),\quad N(t)=n,
 \end{equation}
 where $\kappa_0$ represents an effective ``spring constant'' for an RTP in a harmonic potential, whereas it corresponds to a protein degradation rate in the case of a gene network. Comparison with Eq. (\ref{PDMP}) implies that $A_n(x)=-\kappa_0 x+v_n$. One major difference between an RTP and a gene network is that the continuous variable $X(t)$ has to be positive in the latter case. However, we will assume that the effective ``harmonic potential'' for $v_0>v_1\geq 0$ restricts $X(t)$ to positive values with high probability so that we do not have to impose the condition $X(t)\geq 0$ explicitly. (If $D=0$ then $X(t)\in \Sigma = [v_0/\kappa_0,v_1/\kappa_0]$ and the CK equation can be restricted to the finite interval $\Sigma$ with reflecting boundary conditions at the ends. In this case, the steady-state CK equation can be solved explicitly \cite{Kepler01,Bose04,Smiley10}.)

\subsection{Bare propagator}

First suppose that there is no switching ($\alpha=\beta=0)$. The FP equation for the bare propagator $p_n$ is
\begin{align}
 \frac{\partial p_{n}}{\partial t}&=\frac{\partial (\kappa_0x-v_n)p_{n}}{\partial x}+D\frac{\partial^2p_{n}}{\partial x^2}.
 \label{DLa}
\end{align}
One way to determine the propagator $p_n(x,t|x_0,0)$ is to use the fact that we have a Gaussian process so we only need to determine the first and second moments of $X(t)$. 
Taking expectations of both sides of Eq. (\ref{PDMPrtp}) and using $\langle \d W(t)\rangle =0$ yields the deterministic differential equation
 \[\frac{d\langle X\rangle}{dt}=-\kappa_0\langle X\rangle+v_n.\]
This has the solution 
 \begin{equation}
 \label{mean}
 \langle X(t)\rangle\equiv m_{n}^{(1)}(x_0,t)=x_0\e^{-\kappa_0t} +\frac{v_n}{\kappa_0}(1-\e^{-\kappa_0 t}).
 \end{equation}
  Similarly, using $\langle dX(t) dW(t)\rangle =0$ and $dW(t)^2=dt$, we have
 \begin{align*}
 \langle X(t+dt)^2\rangle &=\langle [X(t)+dX(t) ]^2\rangle\\
 &=\langle [(1-\kappa_0dt)X(t)+vdt+\sqrt{2D}dW(t)]^2\rangle\\
 &=(1-\kappa_0d t )^2\langle X(t)X(t)\rangle\nonumber 
 \\&\quad + v_ndt^2  +2v_n(1-\kappa_0dt)X(t)dt+2Dd t.
\end{align*}
Subtracting $\langle X(t)X(t)\rangle$ from both sides, dividing through by $d t$ and taking the limit $d t \rightarrow 0$ leads to the second-order moment 
equation
\begin{equation}\frac{d\langle X^2\rangle}{dt}=-2\kappa_0 \langle X^2\rangle+2v_n\langle X \rangle +2D,\end{equation}
 which has the solution
 \begin{align}
 \label{var}
 \langle X(t)^2 \rangle &=\e^{-2\kappa_0t}x_0^2 +\frac{D}{\kappa_0}\left (1-\e^{-2\kappa_0t}\right ) \\
 &\quad +\frac{2v_nx_0}{\kappa_0}\e^{-\kappa_0 t}(1-\e^{-\kappa_0t})+\left (\frac{v}{\kappa_0}\right )^2 [1-\e^{-\kappa_0t}]^2.\nonumber
 \end{align}
 It immediately follows that
 \begin{equation}
 \label{eq2:OUvar}
 \mbox{Var}[X(t)]=\frac{D}{\kappa_0}\left (1-\e^{-2\kappa_0t}\right ).
 \end{equation}
Hence, the bare propagator $p_{n}$ has  the explicit solution 
\begin{align}
\label{pdfOU2}
 & p_{n}(x,t|x_0,0) \\
 &=\frac{1}{\sqrt{2\pi \Sigma(t)}} \exp\left (-\frac{[x-x_0\e^{-\kappa_0t}-v_n(1-\e^{-\kappa_0t})/\kappa_0]^2}{2\Sigma(t)}\right ), \nonumber 
\end{align}
\begin{equation}
\Sigma(t)=\frac{D}{\kappa_0}(1-\e^{-2\kappa_0t}).
\end{equation}
\medskip

\subsection{Conditional moments}
When switching is included, the moments of the hSDE are given by the full propagator:
\begin{equation}
M_{nn_0}^{(k)}(x_0,t)=\int_{-\infty}^{\infty}dx\,  x^k G_{nn_0}(x,t|x_0,0).
\end{equation}
Setting
 \begin{align}
 M_{nn_0}^{(k,\mu)}(x_0,t)&=\int_{-\infty}^{\infty}dx\,  x^k   G^{(\mu)}_{nn_0}(x,t|x_0,0), \end{align}
we define the conditional moments according to
 \begin{align}
{\mathcal M}^{(k,\mu)}_{nn_0}(x_0,t)&=\int_{-\infty}^{\infty} dx \, x^k\calG^{(\mu)}_{nn_0}(x,t|x_0,0)\nonumber \\ &=\frac{M_{nn_0}^{(k,\mu)}(x_0,t)}{\sum_{l=0,1}M_{ln_0}^{(0,\mu)}(x_0,t)} ,
\label{calM}
\end{align}
with $\calG^{(\mu)}$ defined in Eq. (\ref{rho}).

\begin{figure*}[t!]
\centering
\includegraphics[width=17cm]{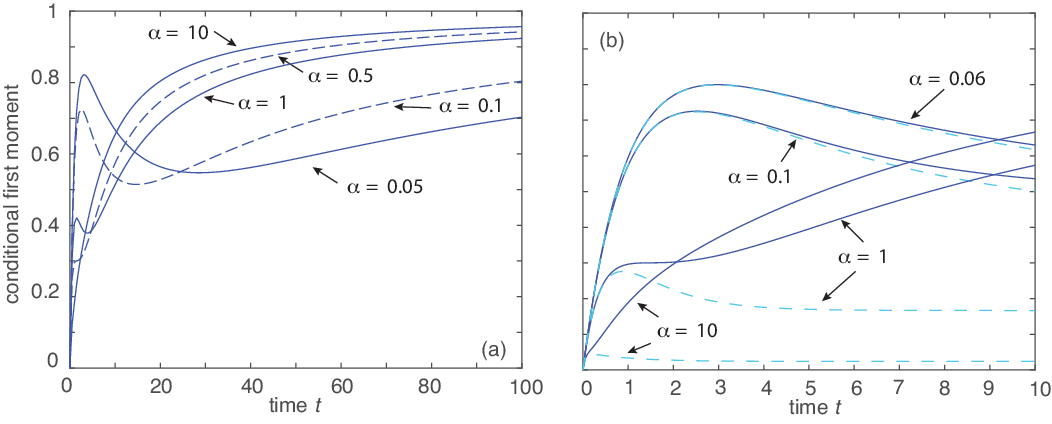} 
\caption{(a) Plot of conditional first moment ${\mathcal M}_{00}^{(1,2)}(x_0,t)$ given by Eq. (\ref{MM}) as a function of time for different values of $\alpha$. Other parameter values are $v_0=-v_1=1$, $\kappa_0=1$ and $x_0=1$. Dashed curve is the bare first moment. The curves approach unity as $t\rightarrow \infty$. (b) Comparison of ${\mathcal M}_{00}^{(1,2)}(x_0,t)$ (solid curves) with the unconditional first moment $M_{00}^{(1)}(x_0,t)$ (dashed curves) given by Eq. (\ref{exact}).}
\label{fig2}
\end{figure*}
 
 For the sake of illustration, consider the first moment for $n=n_0=0$ and $\mu= 2$:
 \begin{align}
{\mathcal M}^{(1,2)}_{00}(x_0,t)&=\int_{-\infty}^{\infty} dx \, x^k\calG^{(2)}_{00}(x,t|x_0,0),\\
&=\frac{M_{00}^{(1,2)}(x_0,t)}{M_{00}^{(0,2)}(x_0,t)+M_{10}^{(0,2)}(x_0,t)} ,
\label{calMex}
\end{align}
 Using Eq. (\ref{tr2}) we have
 \begin{align}
 &M_{00}^{(0,2)}(x_0,t)=\e^{-\beta t}+P_{00}^{(2)}(t) \\
 &=\e^{-\beta t}+\alpha \beta \int_{0}^td\tau_2 \int_{0}^{\tau_2}d\tau_1  \e^{-\beta (t-\tau_2)}   \e^{-\alpha(\tau_2-\tau_1)}\e^{-\beta \tau_1 }\nonumber  \\
 &= \e^{-\beta t}+\frac{\alpha \beta}{\beta-\alpha}\e^{-\beta t}\left [\frac{e^{(\beta-\alpha)t}-1}{\beta-\alpha }-t\right ].\nonumber 
 \end{align}
 In the limit $\beta \rightarrow \alpha$ this reduces to 
 \begin{align}
 \label{moon}
 M_{00}^{(0,2)}(x_0,t)&=\e^{-\alpha t}\left [1+\frac{\alpha^2t^2}{2}\right ].
 \end{align}
 Similarly,
  \begin{align}
  \label{moon1}
 &M_{10}^{(0,2)}(x_0,t)=P_{10}^{(1)}(t) =\frac{\beta}{\alpha-\beta} \left [1-\e^{-(\alpha-\beta)t}\right ]\nonumber\\
 &\rightarrow \alpha t\e^{-\alpha t}\quad \mbox{as } \beta \rightarrow \alpha.
 \end{align}

The corresponding expressions for the full zeroth moments are \cite{Garcia21}.
 \begin{equation}
 M_{00}^{(0)}(x_0,t)=\e^{-\alpha t}\cosh(\alpha t) ,\  M_{10}^{(0)}(x_0,t)=\e^{-\alpha t}\sinh(\alpha t).
 \end{equation}
 Turning to the numerator in Eq. (\ref{calMex})
we find that for $\alpha=\beta$ (see appendix A)
\begin{widetext}
\begin{align}
 M_{00}^{(1,2)}(x_0,t)&= x_0\left (1+\frac{\alpha^2t^2}{2}\right )\e^{-[\kappa_0+\alpha ]t}+\frac{v_0}{\kappa_0} \left (1+\frac{\alpha^2t^2}{2}\right )\left (1-\e^{-\kappa_0t}\right )\e^{-\alpha t} +\frac{2\alpha^2(v_0-v_1)}{\kappa_0^3}\left (1-\e^{-\kappa_0t}\right )\e^{-\alpha t}\nonumber \\
 &\quad -\frac{\alpha^2 t(v_0-v_1)}{\kappa_0^2}\left (1+\e^{-\kappa_0t}\right )\e^{-\alpha t}. 
 \label{ye}
\end{align}
\end{widetext}
If $v_0=-v_1=v$, then our result is consistent with Taylor expanding the exact expression \cite{Garcia21}, which can be written in the form
\begin{align}
 M_{00}^{(1)}(x_0,t)&= x_0\e^{-\kappa_0t}\cosh(\alpha t) \nonumber \\
 &\quad +\frac{v\kappa_0(1-\e^{-\kappa_0t})\cosh(\alpha t)\e^{-\alpha t} }{\kappa_0^2-4\alpha^2}\nonumber \\
 &\quad -\frac{2v\alpha (1+\e^{-\kappa_0t})\sinh(\alpha t)\e^{-\alpha t} }{\kappa_0^2-4\alpha^2}.
\label{exact}
\end{align}
(All $\alpha$-dependent terms are expanded except for the exponential factors $\e^{-\alpha t}$.) Note that Eqs. (\ref{moon}), (\ref{moon1}) and (\ref{ye}) involve the secular terms $\alpha t$ and $(\alpha t)^2$. Hence, $M_{00}^{(1,2)}$ does not yield a good approximation of $M_{00}^{(1)}$ unless $t\ll 1/\alpha$. Similarly for the zeroth moments.

Finally, substituting Eqs. (\ref{moon}), (\ref{moon1}) and (\ref{ye}) into Eq. (\ref{calMex}) yields the following expression for the conditional first moment given a maximum of two transitions:
\begin{align}
\label{MM}
&{\mathcal M}^{(1,2)}_{00}(x_0,t)\nonumber\\
&=\frac{1+\alpha^2 t/2}{1+\alpha t +\alpha^2 t^2/2}\left [x_0 \e^{-\kappa_0t}+\frac{v_0}{\kappa_0}  \left (1-\e^{-\kappa_0t}\right )\right ]\nonumber\\
 &\quad  +\frac{\alpha^2}{1+\alpha t+\alpha^2 t^2/2}\frac{2(v_0-v_1)}{\kappa_0^3 }\left (1-\e^{-\kappa_0t}\right ) \nonumber \\
 &\quad  -\frac{\alpha^2t}{1+\alpha t+\alpha^2 t^2/2}\frac{ (v_0-v_1)}{\kappa_0^2}\left (1+\e^{-\kappa_0t}\right ) .
\end{align}
Note that,
\begin{equation}
\lim_{t\rightarrow \infty} {\mathcal M}^{(1,2)}_{00}(x_0,t)=\frac{v_0}{\kappa_0}.
\end{equation}
The fact that this limit is independent of the leftward velocity $v_1$ reflects the fact that restricting the dynamics to two switching events means that the fraction of time spent in the right-moving state approaches unity in the limit $t\rightarrow \infty$. Note, however, the conditional and bare moments differ significantly for finite $t$. In particular, the conditional moment takes much longer to approach the steady-state, and tends to be a non-monotonic function of $t$, as illustrated in Fig. \ref{fig2}(a). 
In Fig. \ref{fig2}(b), we compare the conditional moment ${\mathcal M}_{00}^{(1,2)}(t)$ with the unconditional moment $M_{00}^{(1)}$ given by Eq. (\ref{exact}) for $v_0=-v_1=v$. The latter has the asymptotic limit
\begin{equation}
\lim_{t\rightarrow \infty} M^{(1)}_{00}(x_0,t)=\frac{v}{\kappa_0+2\alpha}.
\end{equation}
As expected, the difference between the two moments increases with $\alpha$.

\setcounter{equation}{0}
\section{Truncated search process with stochastic resetting} 

\begin{figure}[h!]
\centering
\includegraphics[width=6cm]{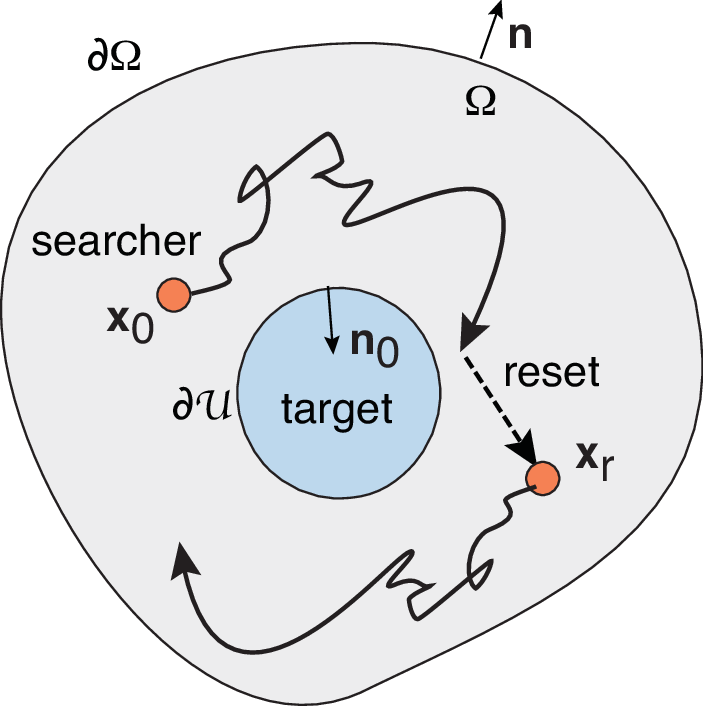} 
\caption{Domain $\Omega \subset \R^d$ containing a single target $\calU$ with a totally absorbing surface $\partial \calU$. Particle starts at $\x_0$ and resets to the point $\x_r$ at a constant rate $r$.}
\label{fig3}
\end{figure}

We now turn to another example of a randomly switching process, namely, a search process with stochastic resetting  \cite{Evans20}. Consider a particle (searcher) subject to Brownian motion in $ \Omega \subseteq \R^d$, and resetting to a fixed point $\x_r$ at a constant rate $r$. Suppose that there exists some target ${\mathcal U} \subset \Omega$ whose boundary $\partial {\mathcal U} $ is totally absorbing and $\x_r\notin {\mathcal U}$, see Fig. \ref{fig3}.
The probability density $p_r(\x,t|\x_0)$ for the particle to be at position $\x$ at time $t$  given the initial position $\x_0$  evolves according to the master equation 
\begin{align}
 \label{reset2}
\frac{\partial p_r(\x,t|\x_0)}{\partial t}&={\bm {\bm \nabla}}^2 p_r(\x,t|\x_0)-rp_r(\x,t|\x_0)\nonumber \\
&\quad +rQ_r(\x_0,t)\delta(\x-\x_r),
\end{align}
where
$Q_r(\x_0,t)$ is the survival probability of a particle that started at $\x_0$:
\begin{equation}
\label{eq7:surv}
Q_r(\x_0,t)=\int_{\Omega\backslash{\mathcal U}} p_r(\x,t|\x_0)d\x.
\end{equation} 
Eq. (\ref{reset2}) is supplemented by the absorbing boundary condition $p_r(\x,t|\x_0)=0$ for all $\x\in \partial {\mathcal U} $ and the reflecting boundary condition $J_r(\x,t|\x_0)=0$ for all $\x\in \partial \Omega$. Here $J_r(\x,t|\x_0)=-{\bm \nabla} p_r(\x,t|\x_0)\cdot \n$ with $\n$ the outward normal on $\partial \Omega$.
Let 
Integrating Eq. (\ref{reset2}) with respect to $\x\in \Omega \backslash \calU$ and using the divergence theorem shows that
\begin{align}
\frac{\partial Q_{r}(\x_0,t)}{\partial t}&=\int_{\partial \calU}{\bm \nabla} p_{r}(\x,t|\x_0)\cdot \n_0 d\x\equiv -J_r(\x_0,t),
 \label{qell0}
\end{align}
with $\n_0$ the normal into $\calU$, see Fig. \ref{fig3}.
Let $\calT$ denote the FPT for absorption at $\partial \calU$. The MFPT can be expressed in terms of $Q_r$ according to
\begin{align}
T_r(\x_0)&\equiv \E[\calT]=- \int_0^{\infty}t \frac{dQ_r(\x_0,t)}{dt}d\tau\nonumber\\
&=\int_0^{\infty}Q_r(\x_0,t)dt.
\label{eq7:TQ}
\end{align}
 We have used the fact that the FPT density is $f_r(\x_0,t)=-  dQ_r(\x_0,t/dt$.

It is well known that $Q_r$ is related to the survival probability without resetting, $Q_0$, according to a last renewal equation \cite{Evans11a,Evans11b,Evans20}:
\begin{align}
\label{renQ}
Q_r(\x_0,t)&=\e^{-rt}Q_0(\x_0,t) \\
&\quad +r\int_0^t\e^{-r[t-\tau]}Q_0(\x_r,t-\tau)Q_r(\x_0,\tau)d\tau.\nonumber
\end{align}
The first term on the right-hand side represents trajectories with no resettings. The integrand in the second term is the contribution from trajectories that last reset at time $\tau\in (0,t)$, and consists of the product of the survival probability starting from $\x_0$ with resetting up to time $t-\tau$ and the survival probability starting from $\x_r$ without any resetting for the time interval $\tau$. Eq. (\ref{renQ}) is the natural analog of the integral Eq. (\ref{renewal0}). The standard method for solving the renewal Eq. (\ref{renQ}) is to introduce the Laplace transform
\begin{equation}
 \widetilde{Q}_r(\x_0,s)=\int_0^{\infty}Q_r(\x_0,t)\e^{-st}dt,
 \end{equation}
and use the convolution theorem. Thus,
Laplace transforming Eq. (\ref{renQ}) and rearranging shows that
\begin{equation}
\label{Qr}
{ \widetilde{Q}_r(\x_0,s)=\frac{ \widetilde{Q}_0(\x_0,r+s)}{1-r \widetilde{Q}_0(\x_r,r+s)}.}
 \end{equation}
The MFPT to reach the target is then given by
\begin{equation}
\label{eq7:Tr}
T_r(\x_0) =\widetilde{Q}_r(\x_0,0)=\frac{ \widetilde{Q}_0(\x_0,r)}{1-r \widetilde{Q}_0(\x_r,r)}.
 \end{equation}

\subsection{Splitting probabilities and conditional MFPTs} 

Following our analysis of hybrid SDEs, we now consider a truncated version of the search process, in which the maximum number of resets is fixed. This is equivalent to truncating the Volterra series expansion of the renewal equation, which in the time domain takes the form
\begin{widetext}
\begin{align}
\label{VolQ}
Q_r(\x_0,t)&=\e^{-rt}Q_0(\x_0,t) \\
&\quad +r\e^{-rt}\int_0^tQ_0(\x_r,\tau)Q_0(\x_r,t-\tau)d\tau+r^2\e^{-rt}\int_0^td\tau_2\int_0^{\tau_2}d\tau_1Q_0(\x_r0\tau_2)Q_0(\x_r,\tau_1)Q_0(\x_r,t-\tau_1-\tau_2)+\ldots. \nonumber
\end{align}
\end{widetext}
The corresponding expansion in Laplace space is a geometric series in powers of  $r\widetilde{Q}_0(\x_r,r+s)$.
The $\ell$th term in the series (\ref{VolQ}), $\ell\geq 0$, is the joint probability $Q_{r,\ell}(\x_0,t)$ that the particle hasn't been absorbed and has reset exactly $\ell$ times:
\begin{align}
&Q_{r,\ell}(\x_0,t)\\
&\quad =r^{\ell}\e^{-rt}[ Q_0(\x_0,\cdot)\otimes Q_0(\x_r,\cdot) \overset{\ell}\otimes Q_0(\x_r,\cdot)](t).\nonumber
\end{align}
where $Q_0\overset{\ell}\otimes Q_0$ denotes the $\ell$th order convolution.
The probability that there are $\ell$ reset events in the interval $[0,t]$ is given by the Poisson distribution
\begin{equation}
P_{\ell}(t)=\frac{(rt)^{\ell}\e^{-rt}}{\ell !}.
\end{equation}
Hence, $Q_{r,\ell}(\x_0,t)/P_{\ell}(t)$ is the survival probability conditioned on exactly $\ell$ reset events in $[0,t]$. In Ref. \cite{Sunil23} the joint probability distribution for the number of resets, the time of absorption, and a general cost was calculated. One result from that analysis was the probability distribution $P(N|\x_0)$ for $N$ resets up to the time of absorption with $\x_r=\x_0$. In our notation, 
\begin{align}
\label{su}
&P(N|\x_0)\\
&=\int_0^{\infty}\left [ \int_0^t \e^{-r(t-\tau)}f_0(t-\tau,\x_0)Q_{r,N}(\tau,\x_0)d\tau\right ]dt.\nonumber
\end{align}
This equation can be interpreted as follows. First, we suppose that the $N$th reset occurred at time $\tau$ and the particle has not yet been absorbed, which is given by the probability $Q_{r,N}(\tau,\x_0)$. The probability density that there are no more resets and the particle is absorbed at time $t$ is then $ \e^{-r(t-\tau)}f_0(t-\tau,\x_0)$. Integrating with respect to $\tau$ and $t$ then yields $P(N|\x_0)$. We can rewrite the right-hand side of Eq. (\ref{su}) using Laplace transforms so that
\begin{align}
\label{su2}
P(N|\x_0)&=\widetilde{f}_0(r,\x_0)\widetilde{Q}_{r,N}(0,\x_0)=\widetilde{f}_0(r,\x_0)[r\widetilde{Q}_{0}(r,\x_0)]^N,
\end{align}
which recovers the result obtained in Ref. \cite{Sunil23}.

In contrast to \cite{Sunil23}, we assume that Brownian motion is killed when either (a) the particle reaches $\partial \calU$ or (b) it resets for the $(\mu+1)$-th times. The unconditional FPT density is then
\begin{equation}
f_{r}^{(\mu)}(\x_0,t)=-\frac{dQ_{r}^{(\mu)}(\x_0,t)}{dt},
\end{equation}
where $Q_{r}^{(\mu)}(x_0,t)$ is the corresponding survival probability:
\begin{equation}
Q_{r}^{(\mu)}(\x_0,t)=\sum_{\ell=0}^{\mu} Q_{r,\ell}(\x_0,t).
\end{equation}
Since $Q_{r}^{(\mu)}(\x_r,0)=1$ and $\lim_{t\rightarrow \infty}Q_{r}^{(\mu)}(\x_r,t)=0$, the FPT density has unit normalization. Using similar arguments to previous examples, the unconditional MFPT is
\begin{align}
T_{r}^{(\mu)}(\x_0)&=\widetilde{Q}_{r}^{(\mu)}(\x_0,0)=\widetilde{Q}_{0}(\x_0,r)\sum_{\ell= 0}^{\mu} \left (r\widetilde{Q}_{0}(\x_r,r)\right )^{\ell}\nonumber \\
&=\widetilde{Q}_0(\x_0,r)\frac{ 1-\left (r \widetilde{Q}_0(\x_r,r)\right )^{\mu+1}}{1-r \widetilde{Q}_0(\x_r,r)}.
\label{conT}
\end{align}

If we wish to distinguish between the two types of killing events, then we need to determine the splitting probabilities and conditional MFPTs. Let $p_{r,\ell}(\x,t|\x_0)$ denote the joint probability density for particle position at time $t$ and the number $\ell$ of resets in the interval $[0,t]$. The forward equation for $p_{r,\ell}$ is
\begin{align}
\frac{\partial p_{r,\ell}(\x,t|\x_0)}{\partial t}&={\bm \nabla}^2 p_{r,\ell}(\x,t|\x_0)-rp_{r,\ell}(\x,t|\x_0)\nonumber \\
&\quad +r\delta(\x-\x_r)Q_{r,\ell-1}(\x_0,t).
 \label{pell}
\end{align}
Integrating with respect to $\x\in \Omega\backslash \calU$ implies that
\begin{align}
\frac{\partial Q_{r,\ell}(\x_0,t)}{\partial t}&=\int_{\partial \calU}{\bm \nabla} p_{r,\ell}(\x,t|\x_0)\cdot \n_0 d\x -rQ_{r,\ell}(\x_0,t)\nonumber \\
&\quad +rQ_{r,\ell-1}(\x_0,t)\nonumber \\
&\equiv -J_{a,\ell}(\x_0,t)- J_{b,\ell}(\x_0,t),
 \label{qell}
\end{align}
with $Q_{r,-1}\equiv 0$.
Here $J_{a,\ell}(\x_0,t)$ is the probability flux into the surface $\partial \calU$,
\begin{equation}
\label{Jell}
J_{a,\ell}(\x_0,t)=-\int_{\partial \calU}{\bm \nabla} p_{r,\ell}(\x,t|\x_0)\cdot \n_0 d\x,
\end{equation}
  whereas $J_{b,\ell}(\x_0,t)$ is the probability flux associated with resetting,
  \begin{equation}
  \label{Jb}
  J_{b,\ell}(\x_0,t)=rQ_{r,\ell}(\x_0,t)-rQ_{r,\ell-1}(\x_0,t).
  \end{equation}

Let $\pi_a^{(\mu)}(\x_0)$ and $\pi_b^{(\mu)}(\x_0)$ denote, respectively, the splitting probabilities for absorption at $\calU$ and resetting for the $(\mu+1)$-th time. Then
\begin{align}
\label{pia}
\pi_a^{(\mu)}(\x_0)&=\int_0^{\infty} \left [\sum_{\ell=0}^{\mu}J_{a,\ell}(\x_0,t)\right ]dt\nonumber \\
&=\lim_{s\rightarrow 0} \sum_{\ell=0}^{\mu}\widetilde{J}_{a,\ell}(\x_0,s),
\end{align}
and
\begin{align}
\label{pib}
\pi_b^{(\mu)}(\x_0)&=\int_0^{\infty} \left [\sum_{\ell=0}^{\mu}J_{b,\ell}(\x_0,t)\right ]dt 
=r\int_0^{\infty}Q_{r,\mu}(\x_0,t)dt\nonumber \\
&=r \lim_{s\rightarrow 0}\widetilde{Q}_{r,\mu}(\x_0,s).
\end{align} 
In order to determine the Laplace transformed flux $\widetilde{J}_a(\x_0,s)$, we Laplace transform Eq. (\ref{pell}) under the initial condition 
${p}_{r,\ell}(\x,0|\x_0)=\delta(\x-\x_0)\delta_{\ell,0}$. This yields the equation
\begin{align}
&{\bm \nabla}^2\widetilde{p}_{r,\ell}(\x,s|\x_0)-(r+s)\widetilde{p}_{r,\ell}(\x,s|\x_r)\nonumber \\
& \quad =-\delta(\x-\x_0)\delta_{\ell,0}-\delta(\x-\x_r)r\widetilde{Q}_{r,\ell-1}(\x_0,s).
 \label{pellLT}
\end{align}
Introduce the Green's function $G(\x,s|\y)$ with
\begin{align}
&{\bm \nabla}^2G(\x,s|\y)-sG(\x,s|\y) =-\delta(\x-\y),
 \label{GLT}
\end{align}
together with the boundary conditions ${\bm \nabla} G\cdot \n =0$ for all $\x\in \partial \Omega$ and $G(\x,s|\y)=0$ for all $\x\in \partial \calU$.
We can then write the solution for $\widetilde{p}_{r,\ell}$ as
\begin{align}
\widetilde{p}_{r,\ell}(\x,s|\x_r)&= G(\x,s+r|\x_0)\delta_{\ell,0}\nonumber \\
&\quad +rG(\x,s+r|\x_r)\widetilde{Q}_{r,\ell-1}(\x_0,s).
\end{align}
Combining with the Laplace transform of Eq. (\ref{Jell}), we have
\begin{align}
\label{Jell2}
&\widetilde{J}_{a,\ell}(\x_0,s)=-\delta_{\ell,0}\int_{\partial \calU}{\bm \nabla} G(\x,s+r|\x_0) \cdot \n_0 d\x\nonumber \\
&\quad -r\widetilde{Q}_{r,\ell-1}(\x_0,s)\int_{\partial \calU}{\bm \nabla} G(\x,s+r|\x_r)\cdot \n_0 d\x .
\end{align}
Note that $-\int_{\partial \calU}{\bm \nabla} G(\x,s|\x_0) \cdot \n_0 d\x$ can be identified with the Laplace transform of the probability flux into the target in the absence of resetting, which we denote by $\widetilde{J}_0(\x_0,s)$.
Finally, substituting this solution into Eq. (\ref{pia}) gives
\begin{align}
\label{pia2}
\pi_a^{(\mu)}(\x_0)&=\ \lim_{s\rightarrow 0}\widetilde{J}_0(\x_0,r+s)\nonumber \\
&\quad +r\lim_{s\rightarrow 0}\left [\widetilde{J}_0(\x_r,r+s)\sum_{\ell=0}^{\mu-1} \widetilde{Q}_{r,\ell}(\x_0,s)\right ].
\end{align}

Let $\calT_a^{(\mu)}(\x_0)$ be the FPT that the particle is absorbed at $\partial \calU$ having started at $\x_0$. Since there is a nonzero probability that the particle never exits at a point on $\partial \calU$ due to resetting for the $(\mu+1)$th time prior to absorption, it follows that the unconditional MFPT $\E[\calT_a^{(\mu)}(y)]=\infty$. This motivates the introduction of the conditional MFPT
\begin{align}
T_a^{(\mu)}(\x_0)&=\E[\calT_a^{(\mu)}(\x_0)|\calT_a^{(\mu)}(\x_0) < \infty].
\end{align}
The conditional FPT density for absorption is
\begin{equation}
f_a^{(\mu)}(\x_0,t)=\frac{\sum_{\ell=0}^{\mu}J_{a,\ell}(\x_0,t)}{\pi_a^{(\mu)}(\x_0)},
\end{equation}
so that
\begin{align}
T_a^{(\mu)}(\x_0)&=\pi_a^{(\mu)}(\x_0)^{-1}\sum_{\ell=0}^{\mu}\int_0^{\infty} tJ_{a,\ell}(\x_0,t)dt\nonumber\\
&=-\pi_a^{(\mu)}(\x_0)^{-1}\sum_{\ell=0}^{\mu}\lim_{s\rightarrow 0}\partial_s \widetilde{J}_{a,\ell}(\x_0,s).
\label{Ta}
\end{align}
Similarly,
\begin{align}
T_b^{(\mu)}(\x_0)&=-\pi_b^{(\mu)}(\x_0)^{-1}r\lim_{s\rightarrow 0}\partial_s \widetilde{Q}_{r,\mu}(\x_0,s).
\label{Tb}
\end{align}

\subsection{Diffusion on the half-line}

\begin{figure}[t!]
\centering
\includegraphics[width=8cm]{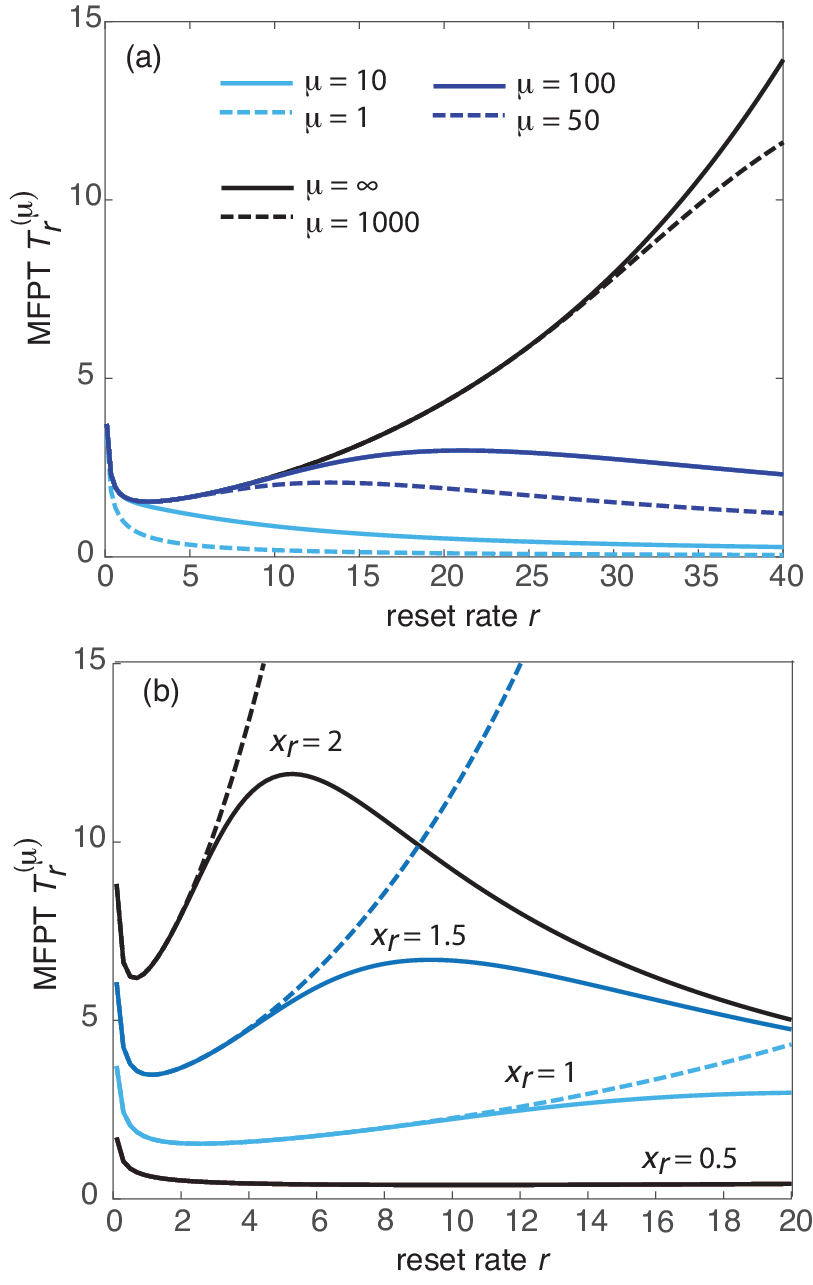} 
\caption{Plot of the MFPT $T_{r}^{(\mu)}(x_r)$ as a function of $r$  for a Brownian particle on the half-line that is killed either by reaching the boundary $x=0$ or by resetting for the $(\mu+1)$-th time. (a) Various $\mu$ for $x_r=1$. (b) Various $x_r$ for $\mu=100$ (solid curves) and $\mu=\infty$ (dashed curves). We set $D=1$.}
\label{fig4}
\end{figure}

\begin{figure*}[t!]
\centering
\includegraphics[width=16cm]{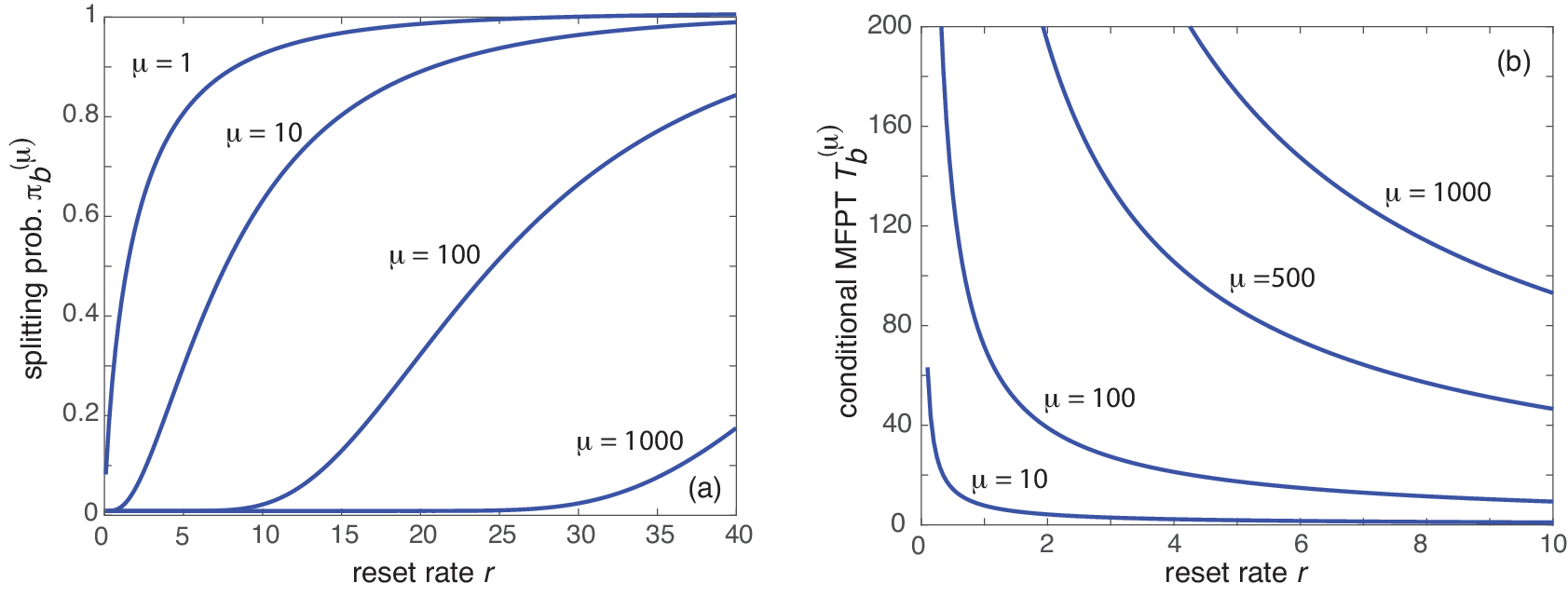} 
\caption{(a) Plot of the splitting probability $\pi_b^{(\mu)}$ as a function of $r$ for the particle to reset for the $(\mu+1)$th time before being absorbed at $\partial \calU$. (b) Corresponding plots of the conditional MFPT $T_b^{(\mu)}$. We set $x_r=1$ and $D=1$.}
\label{fig5}
\end{figure*}

 Consider a diffusing particle on the half-line $[0,\infty)$ with an absorbing target at $x=0$. For simplicity, we set $x_r=x_0$. In the absence of resetting the Laplace transformed survival probability $\widetilde{Q}_0(x,s)$ satisfies the equation
\begin{eqnarray}
D \frac{d^2\widetilde{Q}_0}{dx^2}-s\calQ_0=-1,\quad x\in (0,\infty),
\end{eqnarray}
together with the boundary condition
\begin{eqnarray}
\widetilde{Q}_0(0,s)=0.
\end{eqnarray}
The solution takes the form \cite{Evans11a,Evans11b}
\begin{equation}
\widetilde{Q}_0(x_r,s)=\frac{1-\e^{-\sqrt{s/D}|x_r|}}{s},
\label{eq7:QD}
\end{equation}
which can be inverted to give the error function
\begin{equation}
Q_0(x_r,t)=\mbox{erf}(x_r/2\sqrt{Dt}).
\end{equation}
 Eq. (\ref{eq7:Tr}) then implies that
 \begin{equation}
 \label{eq7:TXr}
T_r (x_r)= \frac{1}{r}\left (\e^{\sqrt{r/D} x_r}-1\right ).
\end{equation}
Note that in the limit $r\rightarrow 0$, the MFPT diverges as 
$T_r \sim 1/\sqrt{r}$, which recovers the result that the MFPT of a Brownian particle without resetting to return to the origin is infinite. One also finds that $T_r$ diverges in the limit $r\rightarrow \infty$, since the particle resets to $x_r$ so often that it never has the chance to reach the origin. Finally, the MFPT has a finite and unique minimum at an intermediate value of the resetting rate $r$ \cite{Evans11b,Evans14}.

If we restrict the maximum number of resets, then the blow-up of $T_r$ at $r\rightarrow \infty$ no longer occurs. This suggests that the unconditional MFPT $T_r^{(\mu)}(x_r)$ may no longer be unimodal. This is indeed found to be the case. In particular, 
Eq. (\ref{conT}) implies that
\begin{align}
&T_{r}^{(\mu)}(x_r)\\
&=\frac{1}{r}\left (\e^{\sqrt{r/D} x_r}-1\right ) \left [ 1-\left (1-\e^{-\sqrt{r/D}|x_r|}\right )^{\mu+1}\right ].\nonumber 
\end{align}
In Fig. \ref{fig4}(a) we plot $T_{r}^{(\mu)}(x_r)$ as a function of the resetting rate $r$ for various values of $\mu$ and fixed $x_r$. For sufficiently small $\mu$, the MFPT is a monotonically decreasing function of $r$, whereas as $\mu$ increases, $T_r^{(\mu)}$ develops a local minimum but is not unimodal. Corresponding plots for various reset positions $x_r$ and fixed $\mu$ are shown in Fig. \ref{fig4}(b). It can be seen that the value of $r$ where truncation starts to have a noticeable effect decreases as $x_r$ increases.

\begin{figure*}[t!]
\centering
\includegraphics[width=16cm]{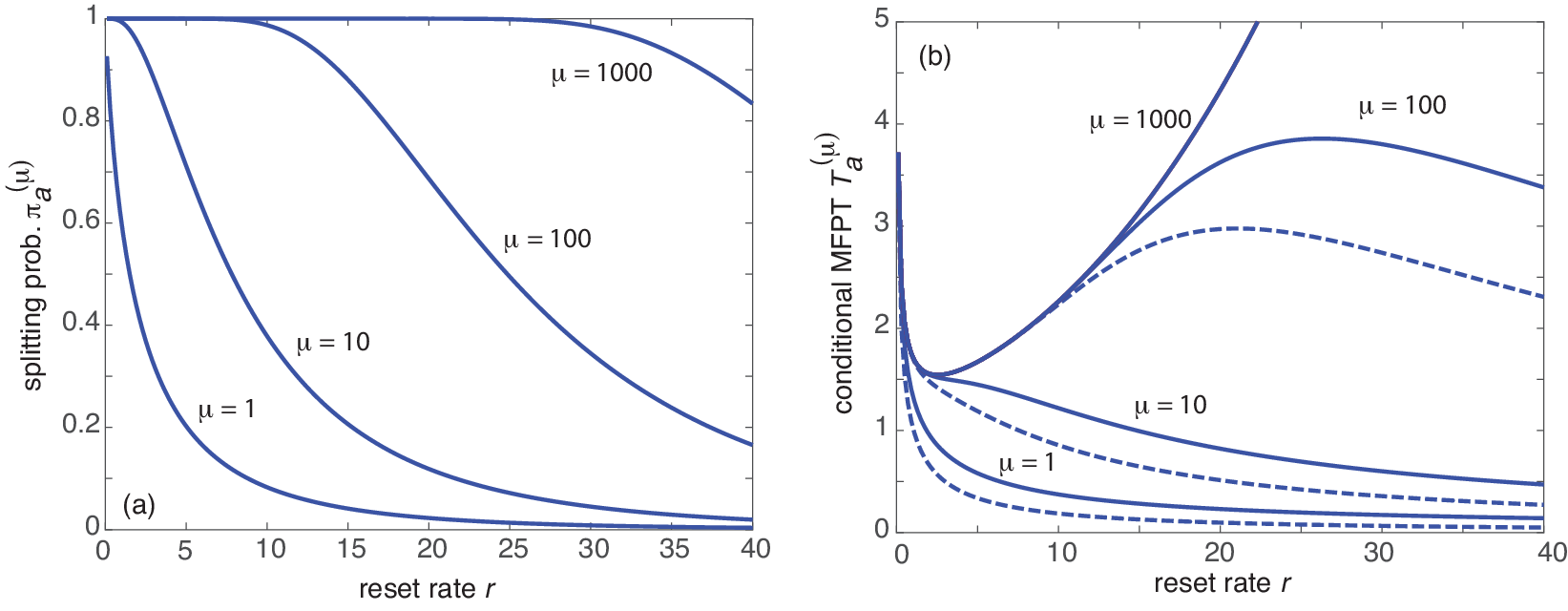} 
\caption{(a) Plot of the splitting probability $\pi_a^{(\mu)}$ as a function of $r$ for the particle to be absorbed at $\partial \calU$ before resetting for the $(\mu+1)$th time. (b) Corresponding plots of the conditional MFPT $T_a^{(\mu)}$ (solid curves) and the unconditional MFPT $T_r^{(\mu)}$ (dashed curves). For $\mu=1000$ the two curves coincide over the given range of $r$. We set $x_r=1$ and $D=1$.}
\label{fig6}
\end{figure*}

Turning to the splitting probabilities and conditional MFPTs, we use the identities $1=\pi_a^{(\mu)}(x_r)+\pi_b^{(\mu)}(x_r)$ and $T_r^{(\mu)}=\pi_a^{(\mu)}T_a^{(\mu)}+\pi_b^{(\mu)}Ti_b^{(\mu)}$. This means that, given $T_r^{(\mu)}$, we only need to calculate $\pi_b^{(\mu)}$ and $T_b^{(\mu)}$.
First, using Eq. (\ref{pib}) we have
\begin{align}
\pi_b^{(\mu)}(x_r)&=
r\widetilde{Q}_{r,\mu}(x_r,0)=\left (r\widetilde{Q}_{0}(x_r,r)\right )^{\mu+1}\nonumber \\
&=\left (1-\e^{-\sqrt{r/D}x_r}\right )^{\mu+1}.
\end{align}
Second, Eq. (\ref{Tb}) becomes
\begin{align}
\pi_b^{(\mu)}(x_r)T_b^{(\mu)}(x_r)&=-r(\mu+1)   \left (r\widetilde{Q}_{0}(x_r,r)\right )^{\mu}\partial_s\widetilde{Q}_{0}(x_r,r)\nonumber \\
&=\frac{\mu+1}{r} \left (1-\e^{-\sqrt{r/D}x_r}\right )^{\mu} \\
&\quad \times \left [1-\frac{[2+x_r\sqrt{r/D}]\e^{-\sqrt{r/D}x_r}}{2} \right ].\nonumber 
\end{align} 
Example plots of $\pi_b^{\mu})$ and $T_b^{(\mu)}$ are shown in Fig. \ref{fig5}.
As expected, the probability $\pi_b^{(\mu)}$ that the particle resets for the $(\mu+1)$th time before being absorbed decreases as the maximum reset threshold $\mu$ is increased. On the other hand, it is an increasing function of $r$. The conditional MFPT for exceeding the reset threshold $\mu$ is a monotonically decreasing function of $r$ and a monotonically increasing function of $\mu$. This is consistent with the idea that, all other things being equal, a faster reset rate reduces the time to reach $\mu+1$. In Fig. \ref{fig6} we show corresponding plots of $\pi_a^{(\mu)}$ and $T_a^{(\mu)}$ with
\begin{equation}
\pi_a^{(\mu)}=1-\pi_b^{(\mu)},\quad T_a^{(\mu)}=\frac{T_r^{(\mu)}-\pi_b^{(\mu)}T_b^{(\mu)}}{\pi_a^{(\mu)}}.
\end{equation}

\section{Discussion} In this paper we explored the effects of restricting the maximum number of switching events in a stochastic hybrid system, under the assumption that switching costs energy. We considered two distinct classes of switching dynamics; (i) an hSDE and (ii) diffusion with stochastic resetting. In the former case, we truncated a Volterra series expansion of the particle propagator, and used this to define a renormalized propagator in which the maximum number of switching events is fixed. We illustrated the theory by calculating the renormalized moments of an OU process with random drift. In case (ii), we truncated a Volterra series expansion of the survival probability of a Brownian particle searching for an absorbing target. This led to a modified FPT problem in which the search is terminated when either the particle finds the target or the number of resets exceeds a fixed threshold. We calculated the splitting probabilities and conditional MFPTs for these mutually exclusive events.

There are a number of natural extensions of the current work. The first is to calculate renormalized propagators for hSDEs beyond the example of a one-dimensional OU process with random drift. One of the challenges is that there are few examples where the bare propagators $p_n$ are known exactly. Moreover, in many cases, the matrix generator ${\bf Q}$ depends on the continuous state $\X(t)$ at time $t$. One notable example is a gene network that is regulated by its own protein product \cite{Kepler01}. Suppose that the promoter has a single operator site $OS_1$ for binding protein X. The gene is assumed to be 
OFF when $X$ is bound to the promoter and 
ON otherwise. If $O_0$ and $O_1$ denote the unbound and bound 
promoter states, then the corresponding state transitions are
$O_0 \overset{\beta x}  \rightarrow O_1$ and $O_1 \overset{\alpha}  \rightarrow O_0$,
where $x$ is the concentration of $X$. Eq. (\ref{PDMPrtp}) still holds but the matrix generator becomes
\begin{equation}
{\bf Q}(x)= \left (\begin{array}{cc} -\beta x  &\quad \alpha 
\\ \beta x& \quad -\alpha  \end{array}
\right ).
\label{autoQ}
\end{equation}
A second example is protein concentration gradient formation during a particular stage of cell polarization in {\em C.~elegans} zygotes. Experimentally, it is found that the underlying mechanism relies on space-dependent switching  between fast and slow diffusion \cite{Wu18}, see also the theoretical studies of Refs. \cite{Bressloff17b,Bressloff19}. Another future direction would be to consider other examples of truncated search processes with stochastic resetting. This could involve modifying the underlying stochastic search dynamics (eg. active particles, L\`evy flights etc.) or introducing delays such as refractory periods and finite return times. Finally, it would be interesting to modify the additive rule for energy cost along the lines of Ref. \cite{Sunil23} by taking the cost of each reset to depend on the distance the particle has to travel to the reset point. This would imply that the threshold $\mu$ for the number of resets before the search process is killed is itself a random variable that depends on the history of previous resets.

 \begin{widetext}

\setcounter{equation}{0}
\renewcommand{\theequation}{A.\arabic{equation}}
\section*{Appendix A: Calculation of truncated first moment $M_{00}^{(1,2)}$}

Consider the truncated first moment of the full propagator $G_{00}(x,t|x_0,0)$, which is defined according to 
\begin{align}
  M_{00}^{(1,2)}(x_0,t)&
  =\int_{-\infty}^{\infty}dx\, x  G^{(2)}_{00}(x,t|x_0,0) . 
 \end{align}
Using Eq. (\ref{tr2}) and the expression for the first moment $ m_{0}^{(1)}(x_0,t)$ of the propagator $p_0$ gives
  \begin{align}
   M_{00}^{(1,2)}(x_0,t)
 &=\e^{-\beta t} m_{0}^{(1)}(x_0,t)+\alpha \beta \int_{0}^td\tau_1 \int_{0}^{\tau_1}d\tau_2 \int_{-\infty}^{\infty} dx_1\int_{-\infty}^{\infty} dx_2 \, \e^{-\beta (t-\tau_1)}m_{0}^{(1)}(x_1,t-\tau_1) \nonumber \\
 &\hspace{3cm} \times  \e^{-\alpha(\tau_1-\tau_2)}p_1(x_1,\tau_1|x_2,\tau_2)\e^{-\beta \tau_2}p_0(x_2,\tau_2|x_0,0)\nonumber \\
  &\quad =\e^{-\beta t} m_{0}^{(1)}(x_0,t) +\alpha\beta \int_{0}^td\tau_1 \int_{0}^{\tau_1}d\tau_2 \int_{-\infty}^{\infty} dx_1\int_{-\infty}^{\infty} dx_2 \, \e^{-\beta (t-\tau_1)} \e^{-\alpha(\tau_1-\tau_2)} \\
 &\quad \quad \times  \left \{\frac{v_0}{\kappa_0}\left (1-\e^{-\kappa_0[t-\tau_1]}\right )+x_1\e^{-\kappa_0[t-\tau_1]}\right \}\e^{-\beta \tau_2}p_1(x_1,\tau_1|x_2,\tau_2)p_0(x_2,\tau_2|x_0,0),\nonumber\end{align}
 after using Eq. (\ref{mean}). Performing the integration with respect to $x_1$ then gives
 \begin{align}
   M_{00}^{(1,2)}(x_0,t) 
  &=\e^{-\beta t} m_{0}^{(1)}(x_0,t)+\alpha \beta \int_{0}^td\tau_1 \int_{0}^{\tau_1}d\tau_2  \int_{-\infty}^{\infty} dx_2 \, \e^{-\beta (t-\tau_1)} \e^{-\alpha(\tau_1-\tau_2)}\e^{-\beta \tau_2}\nonumber \\
  &\quad \times \left [\frac{v_0}{\kappa_0}\left (1-\e^{-\kappa_0[t-\tau_1]}\right )+\left \{\frac{v_1}{\kappa_0}\left (1-\e^{-\kappa_0[\tau_1-\tau_2]}\right )+x_2\e^{-\kappa_0[\tau_1-\tau_2]}\right \} \e^{-\kappa_0[t-\tau_1]}\right ]p_0(x_2,\tau_2|x_0,0).
 \end{align}
Integrating with respect to $x_2$ we have
 \begin{align}
 M_{00}^{(1,2)}(x_0,t)&=\e^{-\beta t} m_{0}^{(1)}(x_0,t)+\alpha \beta \int_{0}^td\tau_1 \int_{0}^{\tau_1}d\tau_2   \, \e^{-\beta (t-\tau_1)} \e^{-\alpha(\tau_1-\tau_2)} \e^{-\beta \tau_2}\\
  &\times  \bigg \{\frac{v_0}{\kappa_0}\left (1-\e^{-\kappa_0[t-\tau_1]}\right )+ \frac{v_1}{\kappa_0}\left (1-\e^{-\kappa_0[\tau_1-\tau_2]}\right )\e^{-\kappa_0[t-\tau_1]} +\left [\frac{v_0}{\kappa_0}\left (1-\e^{-\kappa_0\tau_2}\right )+x_0\e^{-\kappa_0\tau_2}\right ]\e^{-\kappa_0[t-\tau_2]}\bigg \} .  \nonumber \end{align}
Finally, setting $\beta=\alpha$ and computing the time integrals yields Eq. (\ref{ye}).

\end{widetext}

\end{document}